\documentclass[epsf,aps,twocolumn]{revtex4}

\def\be{\begin{equation}}
\def\ee{\end{equation}}
\def\bea{\begin{eqnarray}}
\def\eea{\end{eqnarray}}

\newcommand{\stac}[2]{\stackrel{\scriptscriptstyle {#1}}{#2}}

\begin{document}
%\draft

%<<<<<<<<<<<<< TITLE >>>>>>>>>>>>>>>%
\title{Low energy effective action on a self-gravitating D-brane}

%<<<<<<<<<<<<< AUTHOR >>>>>>>>>>>>>>>%
\author{Sumitada Onda$^{(1)}$, Tetsuya Shiromizu$^{(1,3)}$, Kazuya Koyama$^{(2)}$ 
and Shoko Hayakawa$^{(2)}$}

%<<<<<<<<<<<<< ADDRESS >>>>>>>>>>>>>>>%
\affiliation{$^{(1)}$Department of Physics, Tokyo Institute of Technology, 
Tokyo 152-8551, Japan}

\affiliation{$^{(2)}$Department of Physics, The University of Tokyo, Tokyo 
113-0033, Japan}

\affiliation{$^{(3)}$Advanced Research Institute for Science and Engineering, 
Waseda University, Tokyo 169-8555, Japan}

%<<<<<<<<<<<<< DATE >>>>>>>>>>>>>>>%
\date{\today}

%======================================%
%<<<<<<<<<<<<< ABSTRACT >>>>>>>>>>>>>>>%
%======================================%
\begin{abstract}
Recently the study of braneworld on the self-gravitating D-brane has been initiated 
and 
derived the gravitational equation on the brane by holographic and geometrical 
projection
methods. Surprisingly, in common with these two methods, the matter on the brane 
cannot 
be the source of the gravity on the brane at leading order. In this paper we will 
propose 
the low energy effective action on the D-brane coupled with gravity which derives 
the same results. 
\end{abstract}


\maketitle
%\vskip1cm

%======================================%
%<<<<<<<<<<<< SECTION I  >>>>>>>>>>>>>>%
%======================================%
%\baselineskip25pt
\label{sec:intro}
\section{Introduction}

The purpose of the current paper is the construction of the low energy effective 
action of a self-gravitating D-brane. The action governs the world on the D-brane, %%@
that is, 
D-braneworld. The effect of the self-gravity is essential when one thinks of small 
black holes or very early universe. See Refs. \cite{DBW1,DBW2,DBW3} for cosmology 
related to D-branes. 

Since the D-brane is discovered in superstring theory, properly speaking, the bulk 
spacetimes follows 
supergravity which is the low energy effective theory of superstring theory. The 
brane action is Born-Infeld action. In this line, under $Z_2$-symmetry, the 
gravitational 
equation on the D-brane was derived in Ref. \cite{SKOT}. Since the D-brane carries 
the charge and 
then the gauge field is localized on the brane, we expected the realization of 
Einstein-Maxwell theory on the brane. The result is not the case! At leading order 
the gauge field is 
not the source of the gravity on the brane. 

In this paper, we shall construct the action on the self-gravitating D-brane with 
$Z_2$-symmetry.
Non-trivial issue for the construction is that the gauge field equation exists 
while it does not contribute to the gravity on the brane. The usual expression 
of $\int d^4x {\sqrt {-h}}F^2$ is not permitted, where $h$ is the trace of the 
induced metric 
of the brane $h_{\mu\nu}$. For simplicity, we will work with a toy model \cite{SKT} which is a 
mimic of 
the type IIB supergravity with D-branes. In a toy model, we drop some scalar fields 
because 
we want to concentrate on the behavior of 
gauge fields. We will use the bulk solutions satisfying the junction conditions on 
the brane 
to obtain the final expression for the low energy effective action.  

The rest of this paper is organized as follows. In Sec. II, we describe our toy 
model and 
junction conditions originated from the presence of D-branes. Then we will give the 
bulk 
field solutions satisfying the junction conditions under the long-wave 
approximation \cite{SKOT,SKT,GE,GE2,GE3}. 
The derivation 
will be given in Appendix A. In Sec. III, we first discuss the action for a form 
field 
from the pedagogical point of view and then construct the action for the 
gravitational theory 
on the D-brane where gauge fields is localized. In Appendix B, the 
detail will be described. Finally we will give summary and 
discussion in Sec. 
IV. In Appendix C, from the pedagogical point of view, we discuss the low energy 
effective action 
for the Randall-Sundrum braneworld model \cite{RS,Tess} where the bulk stress 
tensor is just composed of cosmological 
constant and the brane action is assumed that the tension term plus arbitrary four 
dimensional 
matter Lagrangian.

%======================================%
%<<<<<<<<<<<< SECTION II  >>>>>>>>>>>>>%
%======================================%
%\baselineskip25pt
\section{Model}
\label{sec:model}

%--------------------------------------%
%<<<<<<<<<< Subsection A  >>>>>>>>>>>>>%
%--------------------------------------%
\subsection{The action for toy model}

The self-gravitating D-brane was originally considered in type IIB supergravity 
compactified on $S^5$ and it is claimed that the gauge fields which is supposed to 
localise on the brane does not contribute to the source of the gravity on the 
D-brane \cite{SKOT}. The purpose of the current paper is the construction of the low 
energy effective action for that. 

For simplicity, we work with the toy model proposed in Ref. \cite{SKT} 
%===========<Equation>============%
%
\begin{eqnarray}
S  & = & \frac{1}{2\kappa^2} \int_{\cal M} d^5x {\sqrt 
{-g}}\biggl[{}^{(5)}R-2\Lambda
-\frac{1}{2}|H|^2
-\frac{1}{2}(\nabla \chi)^2 \nonumber \\
& & \hskip3cm -\frac{1}{2}|\tilde F|^2-\frac{1}{2}|\tilde
G|^2 \biggr] \nonumber \\
& & + S_{\rm BI} + S_{\rm CS} \nonumber \\
& = & S_{\rm bulk}+ S_{\rm BI} + S_{\rm CS} ,
\label{action}
\end{eqnarray}
%
%=================================%
where $H_{MNK}=\frac{1}{2}\partial_{[M}B_{NK]}$, 
$F_{MNK}=\frac{1}{2}\partial_{[M}C_{NK]}$, 
$G_{K_1 K_2 K_3 K_4 K_5}=\frac{1}{4!}\partial_{[K_1}D_{K_2 K_3 K_4 K_5]}$, 
$\tilde F = F + \chi H$ and $\tilde G=G+C \wedge H$. $M,N,K=0,1,2,3,4$. 
$B_{MN}$ and $C_{MN}$ are 2-form fields, and $D_{K_1 K_2 K_3 K_4}$ is the
4-form field. 
$S_{\rm BI}$ is given by the Born-Infeld action
%===========<Equation>============%
%
\begin{eqnarray}
S_{\rm BI}=-\gamma \int_{\partial {\cal M}} d^4x {\sqrt {-{\rm det}(h+{\cal 
F})}},
\end{eqnarray}
%
%=================================%
where $h_{\mu\nu}$ is the induced metric on the D-brane and 
%===========<Equation>============%
%
\begin{eqnarray}
{\cal F}_{\mu\nu}=B_{\mu\nu}+ \gamma^{-1/2}F_{\mu\nu},
\end{eqnarray}
%
%=================================%
and $F_{\mu\nu}$ is the $U(1)$ gauge field on the brane. $\mu,\nu=0,1,2,3$. 
$S_{\rm CS}$ is Chern-Simons action 
%===========<Equation>============%
%
\begin{eqnarray}
S_{\rm CS} & = & -\gamma \int_{\partial {\cal M}} d^4x {\sqrt {-h}}
\epsilon^{\mu\nu\rho\sigma}\biggl[ \frac{1}{4}{\cal
F}_{\mu\nu}C_{\rho\sigma}+\frac{\chi}{8}{\cal F}_{\mu\nu}{\cal
F}_{\rho\sigma} \nonumber \\
& & \hskip3cm +\frac{1}{24}D_{\mu\nu\rho\sigma} \biggr].
\end{eqnarray}
%
%=================================%
For simplicity we set $H_{\mu\nu\alpha}=0$ and $\tilde F_{\mu\nu\alpha}=0$ in the 
above.
Additionally,
%===========<Equation>============%
%
\begin{eqnarray}
2 \Lambda + \frac{5}{6} \kappa^4 \gamma^2 = 0,
\end{eqnarray}
%
%=================================%
so that the brane geometry could be four dimensional Minkowski spacetime if one 
likes.

The main differences between the original one \cite{SKOT} and the current toy model 
are as follows. In the current model, there are no scalar fields corresponding 
to dilaton and radius of compactification on $S^5$. 
Instead, we introduced the bulk cosmological constant $\Lambda$.
It is turned out that the contribution from scalar fields are not so important when 
one wants to discuss the coupling of the gauge field on the brane to the gravity. 
In fact, we will be able to see the same result, that is, the gauge field localised on 
the brane does not couple to the gravity. The cancellation between the 
contribution from the NS-NS and RR fields is essential.

%--------------------------------------%
%<<<<<<<<<< Subsection B  >>>>>>>>>>>>>%
%--------------------------------------%
\subsection{Junction conditions}

In this subsection we write down the junction conditions which will be used later. 
See appendix A for the bulk field equations. 

Since we are interested in the effective theory on the brane, it is better
for our purpose to adopt the bulk metric 
%===========<Equation>============%
%
\begin{eqnarray}
ds^2=dy^2+g_{\mu\nu}(y,x) dx^\mu dx^\nu,
\end{eqnarray}
%
%=================================%
and  perform (1+4)-decomposition. $y$ is the coordinate orthogonal to the 
brane. 

The junction conditions at the brane located $y=0$ are 
%===========<Equation>============%
%
\begin{eqnarray}
\Bigl[K_{\mu\nu}-h_{\mu\nu}K\Bigr]_{y=0} = 
\frac{\kappa^2}{2}\gamma
(h_{\mu\nu}-T_{\mu\nu}) + {\cal O}(T_{\mu\nu}^2)
\end{eqnarray}
%
%=================================%
%===========<Equation>============%
%
\begin{eqnarray}
H_{y\mu\nu}(0,x)=\kappa^2 \gamma {\cal F}_{\mu\nu}, 
\end{eqnarray} 
%
%=================================%
%===========<Equation>============%
%
\begin{eqnarray}
\tilde F_{y\mu\nu}(0,x)
=\frac{\kappa^2}{2}\gamma \epsilon_{\mu\nu\alpha\beta}{\cal F}^{\alpha\beta},
\end{eqnarray}
%
%=================================%
%===========<Equation>============%
%
\begin{eqnarray}
\tilde G_{y\mu\nu\alpha\beta}(0,x)
=\kappa^2 \gamma \epsilon_{\mu\nu\alpha\beta},
\end{eqnarray}
%
%=================================%
and
%===========<Equation>============%
%
\begin{eqnarray}
\partial_y \chi (0,x) 
= \frac{\kappa^2}{8}\gamma \epsilon^{\mu\nu\alpha\beta}{\cal F}_{\mu\nu}{\cal
F}_{\alpha\beta},
\end{eqnarray}
%
%=================================%
where 
%===========<Equation>============%
%
\begin{eqnarray}
T_{\mu\nu}={\cal F}_{\mu\alpha}{\cal F}^{~\alpha}_\nu -\frac{1}{4}h_{\mu\nu}
{\cal F}_{\alpha\beta} {\cal F}^{\alpha\beta}.
\end{eqnarray}
%
%=================================%
The induced metric on the brane will be denoted by
$h_{\mu\nu}:=g_{\mu\nu}(0,x)$. 

The boundary conditions are specialty for self-gravitating brane which is 
not imposed for probe branes.

\subsection{Solutions at low energy}
\label{sec:approximation}

We solve the bulk field equations with the junction conditions under the long-wave 
approximation(gradient expansion). 

The metric $g_{\mu\nu}(y,x)$ and the extrinsic curvature are expanded as 
%===========<Equation>============%
%
\begin{eqnarray}
g_{\mu\nu}(y,x)
=a^2(y)\Bigl[h_{\mu\nu}(x)+\stac{(1)}{g}_{\mu\nu}(y,x)+\cdots\Bigr].
\end{eqnarray}
%
%=================================%
and
%===========<Equation>============%
%
\begin{eqnarray}
K^{\mu}_{~\nu} = \stac{(0)}{K^{\mu}_{~\nu}}+ 
\stac{(1)}{K^{\mu}_{~\nu}}+\stac{(2)}{K^{\mu}_{~\nu}}+
\cdots.
\end{eqnarray}
%
%=================================%
In the above $\stac{(1)}{g}_{\mu\nu}(0,x)=0 $ and $a(0)=1$. 

As seen in Appendix A, it is easy to obtain the leading order solutions
%===========<Equation>============%
%
\begin{eqnarray}
g_{\mu\nu} =a^2 \Biggl(h_{\mu\nu}+  \frac{3}{8} ( 1 - a^{-16} ) T_{\mu\nu} \Biggr),
\end{eqnarray}
%
%=================================%
%===========<Equation>============%
%
\begin{eqnarray}
K^{\mu}_{~\nu} = - \frac{1}{\ell} \delta^{\mu}_{~\nu} - \frac{1}{2} \kappa^2 \gamma
a^{-16} T^{\mu}_{~\nu},
\label{K}
\end{eqnarray}
%
%=================================%
%===========<Equation>============%
%
\begin{eqnarray}
H_{y\mu\nu}=\kappa^2 \gamma a^{-6}{\cal F}_{\mu\nu},
\label{H}
\end{eqnarray}
%
%=================================%
%===========<Equation>============%
%
\begin{eqnarray}
{\tilde F}_{y\mu\nu}=\frac{1}{2} \kappa^2 \gamma a^{-6} \epsilon_{\mu\nu\alpha\beta} 
{\cal F}^{\alpha\beta},
\label{F}
\end{eqnarray}
%
%=================================%
and
%===========<Equation>============%
%
\begin{eqnarray}
\tilde G_{y\alpha_1 \alpha_2 \alpha_3 \alpha_4}= \kappa^2 
\gamma a^4 \epsilon_{\alpha_1 \alpha_2 \alpha_3 \alpha_4},
\label{G}
\end{eqnarray}
%
%=================================%
where $\epsilon_{\alpha_1 \alpha_2 \alpha_3 \alpha_4}$ is the Levi-Civita 
tensor with respect to the induced metric $h_{\mu\nu}$ on the 
brane. The warp factor is given by
%===========<Equation>============%
%
\begin{eqnarray}
a(y) = e^{- \frac{y}{\ell}},
\end{eqnarray}
%
%=================================%
where
%===========<Equation>============%
%
\begin{eqnarray}
\frac{1}{\ell} = \frac{1}{6}\kappa^2 \gamma.
\end{eqnarray}
%
%=================================%
$\ell$ is the curvature scale of anti-de Sitter like spacetimes. 
This represents the Randall-Sundrum tuning (See Appendix C).
$a(y)$ behaves well for the localization of 
gravity on the brane, that is, we do not encounter a serious 
problem of the localization in our previous work~\cite{SKOT}. However, 
there is the bad behavior of the form fields. Such kind of 
problems will disappear for cases with compactified extra dimensions. The 
compactification radius is supposed to be stabilised somehow. 
Since the compactification naturally introduces the cutoff for the form fields, 
they do not diverge. 

The contribution of $\chi$ field to the action will be higher order, $O({\cal F}^4)$, 
we will omit it hereafter. 

It is useful that the volume element $\sqrt{-g}$ is given by  
%===========<Equation>============%
%
\begin{eqnarray}
\sqrt{-g} =  \sqrt{-h} a^4.
\label{vol}
\end{eqnarray}
%
%=================================%

%======================================%
%<<<<<<<<<<<<< SECTION IV >>>>>>>>>>>>>%
%======================================%
%\baselineskip25pt
\section{Low energy effective action}
\label{sec:action}

Let us consider the low energy effective action on a self-gravitating D-brane.
For the gravitational field $h_{\mu\nu}(x)$, the action has to be derived from 
the five dimensional action by substituting the solutions 
to the bulk field equations 
and integrating out over the bulk coordinate(For example, see Ref. \cite{GE2} or 
Appendix \ref{app}). 
To do so we must take care of the treatment of gauge fields (form fields). 
Generally, for gauge fields, it is known that substituting an ansatz into the 
action 
and the varying of the action does not yield the same result as substituting an ansatz 
into 
the equation of motion if one does not take care of something. 
This problem can be avoided by the addition to the bulk action of total divergence 
terms \cite{ANT, duff, DJ}.
For gauge fields, instead of the Maxwell part of the Born-Infeld action and
the Chern-Simons action, we introduce these divergence terms.

We first give a simple example to see the essence and then construct 
the action for the D-brane. 

\subsection{A simple example}

Let us consider the action of the system which is composed of only $H_{y\mu\nu}$ 
with boundary 
condition on the brane, $H_{y\mu\nu}(0,x)=\kappa^2 \gamma {\cal F}_{\mu\nu}(x)$. 
The fixed background spacetime is anti-deSitter spacetime. For the case with no 
boundary, the action is 
%===========<Equation>============%
%
\begin{eqnarray}
S_0=-\frac{1}{4\kappa^2} \int_{\cal M} d^5x \sqrt{-g} |H|^2.  
\end{eqnarray}
%
%=================================%
But, this is not what we want. Now we have the boundary condition
%===========<Equation>============%
%
\begin{eqnarray}
H_{y\mu\nu}(0,x)=\kappa^2 \gamma {\cal F}_{\mu\nu}(x). \label{bch}
\end{eqnarray}
%
%=================================%
After careful consideration, the action compatible to the boundary 
condition is 
%===========<Equation>============%
%
\begin{eqnarray}
S_1=-\frac{1}{4\kappa^2} \int_{\cal M}d^5x {\sqrt {-g}} \biggl[ |H^2| -  \kappa^2 \gamma 
{\cal F}^{\mu\nu} \partial_y  {\cal F}_{\mu\nu} \biggr],
\end{eqnarray}
%
%=================================%
where ${\cal F}_{\mu\nu}(y,x)=B_{\mu\nu}(y,x)+\gamma^{-1/2}F_{\mu\nu} (x)$. 
If one takes the variation for $B_{\mu\nu}$, we see 
\begin{widetext}
%===========<Equation>============%
%
\begin{eqnarray}
\delta_B S_1 & = & -\frac{1}{4\kappa^2} \int_{\cal M} d^5x {\sqrt {-g}} 
(H^{y\mu\nu} \partial_y \delta B_{\mu\nu}- \kappa^2 \gamma \delta B^{\mu\nu} 
H_{y\mu\nu}
-\kappa^2 \gamma {\cal F}^{\mu\nu}\partial_y \delta B_{\mu\nu}) 
\nonumber \\ 
 & = &  \frac{1}{4\kappa^2} \int_{\cal M} d^5x \delta B_{\mu\nu} \partial_y({\sqrt 
{-g}}H^{y\mu\nu})
 +\frac{1}{2\kappa^2} \int_{\partial {\cal M}}d^4x {\sqrt {-h}} \delta B_{\mu\nu}
(H^{y\mu\nu}-\kappa^2 \gamma {\cal F}^{\mu\nu}). 
\end{eqnarray}
%
%=================================%
Under the boundary condition of Eq. (\ref{bch}), the above reduces 
%===========<Equation>============%
%
\begin{eqnarray}
\delta_B S_1 = \frac{1}{4\kappa^2} \int_{\cal M} d^5x \delta B_{\mu\nu} 
\partial_y({\sqrt {-g}}H^{y\mu\nu})
\end{eqnarray}
%
%=================================%
and then we obtain the bulk field equation for $H_{y\mu\nu}$.

\subsection{Low energy effective action on D-brane}

In similar way with the previous subsection, up to first order, 
the effective action consistent with the junction conditions is given by 
%===========<Equation>============%
%
\begin{eqnarray}
S & = & \frac{1}{2\kappa^2} \int_{\cal M} d^5x {\sqrt {-g}}
\Biggl[ {}^{(5)}R-2\Lambda -\frac{1}{2}|H|^2-\frac{1}{2}|\tilde F|^2 
-\frac{1}{2}|\tilde G|^2+\frac{1}{2}\kappa^2 \gamma {\cal F}^{\mu\nu} \partial_y {\cal F}_{\mu\nu} 
+\frac{1}{4}\kappa^2 \gamma \epsilon^{\mu\nu\alpha\beta}
{\cal F}_{\mu\nu} \partial_y \bigl( C_{\alpha\beta} + \chi {\cal F}_{\alpha \beta} \bigr)
\nonumber \\
& & ~~+\frac{1}{24}\kappa^2 \gamma \epsilon^{\mu\nu \alpha \beta}  \bigl( \partial_y D_{\mu\nu\alpha\beta} 
+ 6 C_{\alpha\beta} \partial_y {\cal F}_{\mu\nu} \bigr)
+N_\mu D_\nu H^{y \mu\nu}\Biggr] - \frac{2}{\kappa^2} \int_{\partial {\cal 
M}} d^4x {\sqrt {-h}} K 
-\gamma \int_{\partial {\cal M}} d^4x {\sqrt {-h}} ,
\label{bulkaction}
\end{eqnarray}
%
%=================================%
where $N^\mu$ is 
the Lagrange multiplier and set to be zero on the brane. The field equation for the gauge field 
is just constraint one as inferred from the bulk field equations in Appendix A. 
The last two terms are the Gibbons-Hawking term and the Nambu-Goto part of the 
Born-Infeld action. 
The factor 2 in the Gibbons-Hawking term comes from the $Z_2$-symmetry of this 
spacetime. Up to first order, the above action is identical to 
%===========<Equation>============%
%
\begin{eqnarray}
S & = & S_{\rm bulk} + \frac{1}{2\kappa^2} \int_{\cal M} d^5x \sqrt{-g} N_{\mu} D_{\nu} H^{y\mu\nu}
+S_{\rm BI}+ S_{\rm CS}-\frac{2}{\kappa^2}\int_{\partial \cal M} d^4x 
{\sqrt {-h}}K. 
\label{boundaryaction}
\end{eqnarray}
%
%=================================%
We should note that an arbitrary tensor 
$d_{\mu \nu \alpha \beta}(x)$, which does not depend on $y$, can be added to 
$D_{\mu \nu \alpha \beta}$ in Eq. (\ref{bulkaction}). This contributes
to the surface integral in Eq. (\ref{boundaryaction}). We should eliminate 
this ambiguity by adding an appropriate boundary term for the form field 
(see Appendix B for the detail). 

Using the equation
%===========<Equation>============%
%
\begin{eqnarray}
{}^{(5)} R = {}^{(4)} R - K^2 - K_{\mu\nu} K^{\mu\nu} - 2 \partial_y K,
\end{eqnarray}
%
%=================================%
together with Eq. (\ref{K}) (See Appendix A), we obtain
%===========<Equation>============%
%
\begin{eqnarray}
{}^{(5)} R - 2 \Lambda = \frac{1}{a^2}{}^{(4)} R(h) - \frac{10}{\ell^2}
\end{eqnarray}
%
%=================================%

Varying this action by each gauge field, indeed, we obtain the correct 
bulk field equations and junction conditions:
%===========<Equation>============%
%
\begin{eqnarray}
\delta_D S= \frac{1}{48\kappa^2} \int_{\cal M} d^5x 
\delta D_{\mu\nu\alpha\beta} \partial_y (\tilde G^{y\mu\nu\alpha\beta}
{\sqrt {-g}})
-\frac{1}{24\kappa^2} \int_{\partial {\cal M}} d^4x {\sqrt {-h}}
\delta D_{\mu\nu\alpha\beta} (\tilde G^{y\mu\nu\alpha\beta} -\kappa^2 \gamma 
\epsilon^{\mu\nu\alpha\beta}),
\end{eqnarray}
%
%=================================%

%===========<Equation>============%
%
\begin{eqnarray}
\delta_B S & = &  \frac{1}{4\kappa^2} \int_{\cal M} d^5x \delta B_{\mu\nu} 
\Biggl[ \biggl( \partial_y \bigl\{ (H^{y\mu\nu}+\chi \tilde F^{y\mu\nu}){\sqrt 
{-g}} \bigr\} 
+\frac{\kappa^2}{2}\gamma \epsilon^{\mu\nu\alpha\beta}
F_{y\alpha\beta} {\sqrt {-g}} \biggr)
\nonumber \\
& & ~~~~~~~~~~~~~~~~~~~~~~~~~~~~~~~~~~~~~
+ \frac{1}{2} \sqrt{-g} \partial_y C_{\alpha\beta} ( \tilde G^{y\mu\nu\alpha\beta} - %%@
\kappa^2 \gamma
\epsilon^{\mu\nu\alpha\beta} ) \Biggr]\nonumber \\
& & -\frac{1}{2\kappa^2} \int_{\partial {\cal M}} d^4x {\sqrt {-h}}
\delta B_{\mu\nu} \biggl[ (H^{y\mu\nu}- \kappa^2 \gamma {\cal F}^{\mu\nu})
+ \chi ( \tilde F^{y\mu\nu} - \frac{\kappa^2}{2} \gamma 
\epsilon^{\mu\nu\alpha\beta} {\cal F}_{\alpha\beta})
\nonumber \\
& & ~~~~~~~~~~~~~~~~~~~~~~~~~~~~~~~~~~~~~
+ \frac{1}{2} C_{\alpha\beta} ( \tilde G^{y\mu\nu\alpha\beta} - \kappa^2 \gamma 
\epsilon^{\mu\nu\alpha\beta} ) \biggr],
\end{eqnarray}
%
%=================================%
%===========<Equation>============%
%
\begin{eqnarray}
\delta_C S & = & \frac{1}{4\kappa^2} \int_{\cal M} d^5x  \delta C_{\mu\nu}
\Biggl[ \biggl\{ \partial_y (\tilde F^{y\mu\nu}{\sqrt {-g}}) - \frac{\kappa^2}{2} 
\gamma 
\epsilon^{\mu\nu}_{~~~\alpha\beta} H^{y\alpha\beta} {\sqrt {-g}} \biggr\}
- \frac{1}{2} H_{y\alpha\beta} ( \tilde G^{y\mu\nu\alpha\beta} - \kappa^2 \gamma 
\epsilon^{\mu\nu\alpha\beta} )
\sqrt{-g} \Biggr]
\nonumber \\
& & -\frac{1}{2\kappa^2} \int_{\partial {\cal M}} d^4 x {\sqrt {-h}} \delta 
C_{\mu\nu} 
(\tilde F^{y\mu\nu} -\frac{\kappa^2}{2}\gamma \epsilon^{\mu\nu\alpha\beta}{\cal 
F}_{\alpha \beta}),
\end{eqnarray}
%
%=================================%
\end{widetext}
respectively. We can confirm that for the variation in terms of each gauge field 
the junction conditions on the brane and the bulk field equations are satisfied. We 
can also derive the gravitational field equation from Eq. (\ref{boundaryaction}) directly. 

Thus we can derive the effective action by substituting the solutions of all fields 
into this action and
integrating out over the coordinate of extra dimensions $y$. Up to first order we 
obtain
%===========<Equation>============%
%
\begin{eqnarray}
S_{\rm eff}=\frac{\ell}{2\kappa^2} \int d^4x{\sqrt {-h}}\Biggl[ {}^{(4)}R(h) 
+\tilde 
N_\mu D_\nu {\cal F}^{\nu\mu}\Biggr]
\end{eqnarray}
%
%=================================%
where $\tilde N^{\mu} \propto N^{\mu}$.
Therefore we see that this action is consistent with the Einstein equation on the 
brane with using the junction conditions and the bulk field equations.

%======================================%
%<<<<<<<<<<<<< SECTION V  >>>>>>>>>>>>>%
%======================================%
%\baselineskip25pt
\section{Summary and discussion}
\label{sec:summary}

In this paper we proposed the low energy effective action for the gravitational 
theory on a toy self-gravitating D-brane model with $Z_2$-symmetry. 
At leading order the gauge field does not couple with the gravity 
on the brane. Yet, 
we can see that the field equations for the gauge fields come from the constraints. 
The Lagrange 
multiplier is set to be zero on the brane. 

There are many remaining issues. First of all, $Z_2$-symmetry is assumed here 
although there is no reason to assume that in 
general for D-brane. It is easy to see that the absence of that symmetry can alter 
the result obtained here. Second, the case of the brane with the net cosmological 
constant. This can be realized by breaking the BPS condition
In this case, the gauge field seems to couple with the gravity on the brane 
\cite{SKT} (see Appendix B). It is 
unlikely that the action is normalisable due to the bad behavior of the gauge field 
in the bulk. 
However, this problem seems to be relaxed if one thinks of the compact extra 
dimensions. 
We hope these issues will be fixed in future study.

%======================================%
%<<<<<<<<< Acknowledgments  >>>>>>>>>>>%
%======================================%
%\baselineskip25pt
\section*{Acknowledgments}

We would like to thank Katsushi Ito, Shinji Mukohyama, Norisuke Sakai and 
Takahiro Tanaka for fruitful discussions. 
SO thanks Akio Hosoya and Masaru Siino for their continuous encouragement.
To complete this work, the discussion during and after the YITP workshops 
YITP-W-01-15 and  YITP-W-02-19
were useful. The work of TS was supported by Grant-in-Aid for Scientific
Research from Ministry of Education, Science, Sports and Culture of 
Japan(No.13135208, No.14740155 and No.14102004). The work of KK and SH was 
supported by JSPS.

\appendix
%======================================%
%<<<<<<<<<<<< APPENDIX A >>>>>>>>>>>>>>%
%======================================%

\section{Basic equations and solutions for toy model}

The ``evolutional" equations to the $y$-direction are 
%===========<Equation>============%
%
\begin{eqnarray}
\partial_y K 
= R-\kappa^2 \biggl( {}^{(5)}T^{\mu}_{~\mu} -\frac{4}{3}{}^{(5)}T^{M}_{~M} \biggr) 
-K^2,
\end{eqnarray}
%
%=================================%
%===========<Equation>============%
%
\begin{eqnarray}
\partial_y \tilde K^{\mu}_{~\nu} = \tilde R^{\mu}_{~\nu} 
-\kappa^2\biggl({}^{(5)}T^{\mu}_{~\nu}
-\frac{1}{4}
\delta^{\mu}_{~\nu} {}^{(5)}T^{\alpha}_{~\alpha} \biggr)-K \tilde K^{\mu}_{~\nu},
\label{traceless}
\end{eqnarray}
%
%=================================%
%===========<Equation>============%
%
\begin{eqnarray}
\partial_y^2 \chi +D^2 \chi +K\partial_y \chi-\frac{1}{2}H_{y\alpha\beta}\tilde 
F^{y\alpha\beta}=0,
\end{eqnarray}
%
%=================================%
%===========<Equation>============%
%
\begin{eqnarray}
\partial_y X^{y\mu\nu}+KX^{y\mu\nu}
+\frac{1}{2}F_{y\alpha\beta}\tilde G^{y\alpha\beta\mu\nu}=0,
\end{eqnarray}
%
%=================================%
%===========<Equation>============%
%
\begin{eqnarray}
\partial_y \tilde F^{y\mu\nu}+K \tilde F^{y\mu\nu}-\frac{1}{2}H_{y\alpha\beta}
\tilde G^{y\alpha\beta\mu\nu}=0,
\end{eqnarray}
%
%=================================%
%===========<Equation>============%
%
\begin{eqnarray}
\partial_y \tilde G_{y \alpha_1 \alpha_2 \alpha_3 \alpha_4}
=K\tilde  G_{y \alpha_1 \alpha_2 \alpha_3 \alpha_4},
\end{eqnarray}
%
%=================================%
where $X^{y\mu\nu}:=H^{y\mu\nu}+\chi \tilde F^{y\mu\nu}$ and the 
energy-momentum tensor is
%===========<Equation>============%
%
\begin{eqnarray}
\kappa^2\;{}^{(5)\!}T_{MN} & = &  \frac{1}{2}\biggl[ \nabla_M \chi \nabla_N \chi
-\frac{1}{2}g_{MN} (\nabla \chi)^2 \biggr]
\nonumber \\
& & +\frac{1}{4}\biggl[H_{MKL}H_N^{~KL}-g_{MN}|H|^2 \biggr] 
\nonumber \\
& &  +\frac{1}{4}\biggl[\tilde F_{MKL}\tilde
F_N^{~KL}-g_{MN}|\tilde F|^2
\biggr]
\nonumber \\
& & 
 +\frac{1}{96}\tilde G_{MK_1 K_2 K_3 K_4} \tilde G_{N}^{~~K_1
K_2 K_3 K_4}
\nonumber \\
& & -\Lambda g_{MN}.
\end{eqnarray}
%
%=================================%
$K_{\mu\nu}$ is the extrinsic curvature, $K_{\mu\nu}=\frac{1}{2}\partial_y 
g_{\mu\nu}$. 
$\tilde K^{\mu}_{~\nu}$ and $\tilde R^{\mu}_{~\nu}$ are the traceless parts
of $K^{\mu}_{~\nu}$ and $R^{\mu}_{~\nu}$, respectively.

The constraints are 
%===========<Equation>============%
%
\begin{eqnarray}
-\frac{1}{2}\biggl[R-\frac{3}{4}K^2+\tilde K^{\mu}_{~\nu} \tilde K^{\nu}_{~\mu} 
\biggr]
=\kappa^2\:{}^{(5)\!}T_{yy},
\end{eqnarray}
%
%=================================%
%===========<Equation>============%
%
\begin{eqnarray}
D_\nu K^{\nu}_{~\mu}-D_\mu K = \kappa^2\:{}^{(5)\!}T_{\mu y},
\end{eqnarray}
%
%=================================%
%===========<Equation>============%
%
\begin{eqnarray}
D^\alpha X_{y\alpha\mu}=0, \label{A10}
\end{eqnarray}
%
%=================================%
%===========<Equation>============%
%
\begin{eqnarray}
D^\alpha \tilde F_{y\alpha\mu}=0,
\end{eqnarray}
%
%=================================%
%===========<Equation>============%
%
\begin{eqnarray}
D^\alpha \tilde G_{y \alpha \mu_1 \mu_2 \mu_3}=0,
\end{eqnarray}
%
%=================================%
where $D_\mu$ is the covariant derivative with respect to $g_{\mu\nu}$.

The first order equations for $\tilde F_{y\mu\nu}$ and $H_{y\mu\nu}$ are 
%===========<Equation>============%
%
\begin{eqnarray}
\partial_y \stac{(1)}{\tilde F}_{y\mu\nu}-\frac{1}{2a^4}
\stac{(1)}{H}_{y\alpha\beta} \tilde 
G_{y\rho\sigma\mu\nu}h^{\alpha\rho}h^{\beta\sigma}=0,
\end{eqnarray}
%
%=================================%
and
%===========<Equation>============%
%
\begin{eqnarray}
\partial_y \stac{(1)}{H}_{y\mu\nu}+\frac{1}{2a^4}\stac{(1)}{\tilde 
F}_{y\alpha\beta}\tilde G_{y\rho\sigma\mu\nu}h^{\alpha\rho}h^{\beta\sigma}=0.
\end{eqnarray}
%
%=================================%

Together with the junction conditions and the solution of $\tilde G_{y\mu_1 \mu_2 
\mu_3 \mu_4}$, these solutions are given by
%===========<Equation>============%
%
\begin{eqnarray}
\stac{(1)}{H}_{y\mu\nu}=\kappa^2 \gamma a^{-6}{\cal F}_{\mu\nu},
\end{eqnarray}
%
%=================================%
and
%===========<Equation>============%
%
\begin{eqnarray}
\stac{(1)}{\tilde F}_{y\mu\nu}=\frac{1}{2} \kappa^2 \gamma 
\epsilon_{\mu\nu\alpha\beta} 
a^{-6}{\cal F}^{\alpha\beta}.
\end{eqnarray}
%
%=================================%
Using these results the "evolutional" equation for the traceless part 
of the extrinsic curvature is 
%===========<Equation>============%
%
\begin{eqnarray}
\partial_y \stac{(1)}{\tilde K^{\mu}_{~\nu}}
= - \stac{(0)} K \stac{(1)}{\tilde K^{\mu}_{~\nu}}
+ a^{-2} {}^{(4)}\tilde R^{\mu}_{~\nu} (h) - \frac{(\kappa^2\gamma)^2}{a^{16}} 
T^{\mu}_{~\nu},
\end{eqnarray}
%
%=================================%
where ${}^{(4)}R^{\mu}_{~\nu} (h)=h^{\mu\alpha}{}^{(4)}R_{\alpha\nu}(h)$ 
is the Ricci tensor with respect to $h_{\mu\nu}$
and $T^{\mu}_{~\nu}= h^{\mu\alpha}T_{\alpha\nu}$. 
The solution is 
%===========<Equation>============%
%
\begin{eqnarray}
\stac{(1)}{\tilde K^{\mu}_{~\nu}}(y,x) & = & -\frac{\ell}{2a^2}{}^{(4)}\tilde 
R^{\mu}_{~\nu} (h) 
- \frac{\kappa^2 \gamma}{2 a^{16}} T^{\mu}_{~\nu} 
+ \frac{\zeta^{\mu}_{~\nu}(x)}{a^4},
\nonumber \\
\end{eqnarray}
%
%=================================%
where $\zeta^{\mu}_{~\nu}$ is the "constant of integration" and expresses
the holographic CFT stress tensor \cite{holo}.
It is a homogeneous solution which satisfies $\zeta^{\mu}_{~\mu}=0$ and $D_{\mu} 
\zeta^{\mu}_{~\nu}=0$.
(This corresponds to the dark radiation at this order.)
Since it is not affect the result below, we will omit $\zeta^{\mu}_{~\nu}$ 
hereafter.

The trace part of the extrinsic curvature can be evaluated 
from the Hamiltonian constraint as 
%===========<Equation>============%
%
\begin{eqnarray}
\stac{(1)}{K}(y,x)=-\frac{\ell}{6a^2} {}^{(4)}R(h).
\end{eqnarray}
%
%=================================%
Finally, we obtain
%===========<Equation>============%
%
\begin{eqnarray}
\bigl( K^{\mu}_{~\nu} - \delta^{\mu}_{~\nu} K \bigr)^{(1)}
= - \frac{\ell}{2 a^2} {}^{(4)}G^{\mu}_{~\nu}(h)
- \frac{\kappa^2 \gamma}{2 a^{16}} T^{\mu}_{~\nu}.
\end{eqnarray}
%
%=================================%

From the junction condition, it is easy to see that the Einstein equation
up to first order becomes
%===========<Equation>============%
%
\begin{eqnarray}
{}^{(4)}G^{\mu}_{~\nu}(h) = 0.
\end{eqnarray}
%
%=================================%
The stress energy tensor of the gauge field does not appear as the source of four 
dimensional 
gravity in this order. Yet, the field equation for the gauge field exists which 
comes from 
the constraint equation for $B_{\mu y}$ in five dimensional sense(Eq. (\ref{A10})). 

Using this solution, up to first order the extrinsic curvature is expressed as 
follows:
%===========<Equation>============%
%
\begin{eqnarray}
K^{\mu}_{~\nu} = - \frac{1}{\ell} \delta^{\mu}_{~\nu} - \frac{1}{2} \kappa^2 \gamma
a^{-16} T^{\mu}_{~\nu}.
\label{K}
\end{eqnarray}
%
%=================================%

\section{Cancellation of U(1) gauge field and BPS condition}

In order to see the cancellation of U(1) gauge field, we will show a 
slightly different derivation of the effective action.  
Integrating the solutions Eqs. (\ref{H}), (\ref{F}) and (\ref{G}), 
we get the solutions for $B_{\mu \nu}, C_{\mu \nu}$ and $D_{\mu \nu \alpha \beta}$

\begin{equation}
B_{\mu \nu}= \frac{\ell}{6} \kappa^2 a^{-6} \gamma {\cal F}_{\mu \nu} +b_{\mu \nu}(x),
\end{equation}
\begin{equation}
C_{\mu \nu} = \frac{\ell}{12} \kappa^2 \gamma 
\epsilon_{\mu \nu \alpha \beta} a^{-6} {\cal F}^{\alpha \beta} +c_{\mu \nu}(x),
\end{equation}
\begin{eqnarray}
D_{\mu \nu \alpha \beta} &=& -\frac{\ell}{4} \kappa^2 \gamma a^{4}
\epsilon_{\mu \nu \alpha \beta} 
-\frac{\ell}{4} \kappa^2 \gamma a^{-12} \epsilon_{\mu \nu \rho \sigma}
{\cal F}^{\rho \sigma} {\cal F}_{\alpha \beta} \nonumber\\
&& - \ell \kappa^2 \gamma {\cal F}_{\mu \nu} c_{\alpha \beta}
 +d_{\mu \nu \alpha \beta}(x),
\end{eqnarray}
where $b_{\mu \nu}$, $c_{\mu \nu}$ and $d_{\mu \nu \alpha \beta}$ are
the constants of integration. By definition, $b_{\mu \nu}$ is given by
\begin{equation}
b_{\mu \nu}(x) = - \gamma^{-1/2} F_{\mu \nu}(x).
\end{equation}

Then the brane action(BI action and CS term) can be evaluated as 
\begin{eqnarray}
S_{\rm CS} &=& \gamma \int d^4 x \sqrt{-h} \epsilon^{\mu \nu \alpha \beta}
\left[ \frac{1}{4} {\cal F}_{\mu \nu} C_{\alpha \beta} +
\frac{1}{24} D_{\mu \nu \alpha \beta} \right] \nonumber\\
& = &  \int d^4 x \sqrt{-h}\left[ -\frac{3}{2} \gamma+ \frac{\gamma}{4}
 {\cal F}_{\mu \nu} {\cal F}^{\mu \nu} \right. \nonumber\\
&&  \left. - \frac{\gamma}{24} \int d^4 x \sqrt{-h}
 \epsilon^{\mu \nu \alpha \beta} d_{\mu \nu \alpha \beta}
\right],
\end{eqnarray}
and 
\begin{equation} 
S_{\rm BI} =  \int d^4 x \sqrt{-h} \left[-\gamma  
- \frac{\gamma}{4}
{\cal F}_{\mu \nu}{\cal F}^{\mu \nu}
\right].
\end{equation}
The bulk action plus GH term $S_{\rm bulk}= S + 2 S_{GH}$ is given by
\begin{equation}
S_{\rm bulk} = \int d^4 x \sqrt{-h}
\left[ \frac{5}{2} \gamma + \frac{\ell}{2 \kappa^2} {}^{(4)}R(h) 
\right].
\end{equation}
Then the total effective action becomes
\begin{equation}
S_{\rm eff}= \frac{\ell}{2 \kappa^2}\int d^4 x \sqrt{-h} {}^{(4)}R(h).
\end{equation}
Here we subtracted the final term in CS action from the total action
by adding a boundary term
\begin{equation}
\frac{\gamma}{24} \int d^4x \sqrt{-h} \epsilon^{\mu \nu \alpha \beta} d_{\mu \nu %%@
\alpha \beta}.
\end{equation}
This is the same boundary term that appears in the derivation of Eq. %%@
(\ref{boundaryaction}) from Eq. (\ref{bulkaction}). 

We can clearly see that, for U(1) gauge field, the contribution from BI term is 
canceled by the contribution from CS term. This is the consequence of the 
BPS condition. If the BPS condition is broken and the coefficients of the 
BI action and CS action are different, the cancellation does not 
occur and the U(1) gauge field appears in the effective action and 
couples to the gravity.  

\section{Low energy effective action for Randall-Sundrum model}
\label{app}

In this section, we derive the low energy effective action on the brane in 
Randall-Sundrum braneworld model. We will use the bulk field solution obtained 
by the long-wave approximation with the junction conditions \cite{GE2}.

It is easy to obtain the zeroth order solutions.
Without derivation we present them.
%===========<Equation>============%
%
\begin{eqnarray}
\stac{(0)}{K^{\mu}_{~\nu}} = - \frac{1}{\ell} \delta^{\mu}_{~\nu},
\end{eqnarray}
%
%=================================%

%===========<Equation>============%
%
\begin{eqnarray}
\stac{(0)}g_{\mu\nu} = a^2(y)h_{\mu\nu}(x) = e^{-\frac{2}{\ell}y}h_{\mu\nu}(x),
\end{eqnarray}
%
%=================================%
where $\ell$ is the curvature radius of $AdS_5$ spacetime.
Here the relations between the tension $\sigma$ and $\ell$ 
%===========<Equation>============%
%
\begin{eqnarray}
\kappa^2 \sigma = \frac{6}{\ell}
\end{eqnarray}
%
%=================================%
have been assumed. It is known as Randall-Sundrum fine tuning.
This tuning means that the effective cosmological constant 
on the brane set zero.

Thus, using the junction condition, up to first order, the extrinsic curvature is
%===========<Equation>============%
%
\begin{eqnarray}
K^{\mu}_{~\nu} (y,x) = -\frac{1}{\ell} \delta^{\mu}_{~\nu} 
-\frac{\kappa^2}{2a^2}\biggl(T^{\mu}_{~\nu} 
-\frac{1}{3}\delta^{\mu}_{~\nu} T \biggr) (h)
\end{eqnarray}
%
%=================================%

We calculate the metric at first order, $\stac{(1)} g_{\mu\nu}$, which will be used 
in the 
computation of the volume element $\sqrt{-g}$.
Together with the junction condition, the result is
%===========<Equation>============%
%
\begin{eqnarray}
\stac{(1)}{g^{\mu}_{~\nu}}(y,x) = \frac{\kappa^2 \ell}{2} (1 - a^{-2})
\biggl( T^{\mu}_{~\nu} - \frac{1}{3} \delta^{\mu}_{~\nu} T \biggr)(h).
\end{eqnarray}
%
%=================================%

Let us consider an effective action for $h_{\mu\nu}(x)$. 
The effective action should be derived
using the solution for the equation of motion with the junction condition.
We shall start with the following action:
\begin{widetext}
%===========<Equation>============%
%
\begin{eqnarray}
S & = & \frac{1}{2 \kappa^2} \int_{\cal M} d^5x \sqrt{-g}
\biggl( {}^{(5)}R + \frac{12}{\ell^2} \biggr)
- \frac{2}{\kappa^2} \int_{\partial \cal M} d^4x \sqrt{-h} K
- \sigma \int_{\partial \cal M} d^4x \sqrt{-h} + \int_{\partial {\cal M}} d^4x 
\sqrt{-h} {\cal L}_{\rm matter},
\end{eqnarray}
%
%=================================%
\end{widetext}
where we have taken into account the boundary term,
the so-called Gibbons-Hawking term, instead of introducing
delta-function singularities in the curvature.
The factor 2 in the Gibbons-Hawking term comes from
the $Z_2$-symmetry of this spacetime.
That is why we can get the effective action on the brane by simple
substitution of the solution for $h_{\mu\nu}$ with using the junction condition.

Further, we can use the equation
%===========<Equation>============%
%
\begin{eqnarray}
{}^{(5)}R & = &  {}^{(4)}R-K^2-K^{\mu\nu}K_{\mu\nu}-2\partial_y K \nonumber \\
          & = & {}^{(4)}R(h)a^{-2}-\frac{20}{\ell^2}+\frac{\kappa^2}{\ell a^2}T.
\end{eqnarray}
%
%=================================%

Therefore, up to first order, we obtain
\begin{widetext}
%===========<Equation>============%
%
\begin{eqnarray}
S & = &  \frac{1}{2\kappa^2}  \int_{\cal M} d^5x {\sqrt {-h}}a^4
\Biggl[1-\frac{\ell^2}{12}(1-a^{-2})\kappa^2 \ell^{-1}T \Biggr]
\Biggl[{}^{(4)}R a^{-2}-\frac{8}{\ell^2}+\frac{\kappa^2}{\ell}\frac{1}{a^2}T 
\Biggr]
+\int_{\partial {\cal M}} d^4x {\sqrt {-h}}\Biggl[-\frac{6}{\ell \kappa^2} + {\cal 
L}_{\rm matter} \Biggr] \nonumber \\
& & -\frac{2}{\kappa^2} \int_{\partial \cal M} d^4 x {\sqrt 
{-h}}\Biggl[-\frac{4}{\ell}+\frac{\kappa^2}{6}T \Biggr] 
\nonumber \\
& = &  \frac{1}{\kappa^2} \int_{\partial \cal M} d^4 x {\sqrt {-h}} 
\Biggl[\frac{\ell}{2}{}^{(4)}R
+ \kappa^2 {\cal L}_{\rm matter} \Biggr]
\end{eqnarray}
%
%=================================%
\end{widetext}
Therefore, up to first order, we find that this effective action is consistent with
the Einstein equation on the brane with using the junction condition.


\begin{thebibliography}{22}

\bibitem{DBW1}
S. Kachru, R. Kallosh, A. Linde, J. Maldacena, L. McAllister and S. P. Trivedi, JCAP {\bf 0310},013(2003);
C. P. Burgess, P. Martineau, F. Quevedo and R. Rabadan, JHEP {\bf 06}, 037(2003);
C. P. Burgess, N. E. Grandi, F. Quevedo and R. Rabadan, hep-th/0310010;
K. Takahashi and K. Ichikawa, hep-th/0310142.  

\bibitem{DBW2}
T. Shiromizu, T. Torii and T. Uesugi, Phys. Rev. {\bf D67}, 123517(2003);
M. Sami, N. Dadhich and T. Shiromizu, Phys. Lett. {\bf B568},118(2003);
E. Elizalde, J. E. Lidsey, S. Nojiri and S. D. Odintsov, hep-th/0307177;
T. Uesugi, T. Shiromizu, T. Torii and K. Takahashi, hep-h/0310059, 
to appear in Phys. Rev. {\bf D}(2004).

\bibitem{DBW3}
S. B. Giddings, S. Kachru and J. Polchinski, Phys. Rev. {\bf D66}, 106006(2002);
O. DeWolfe and S. B. Giddings, Phys. Rev. {\bf D67}, 066008(2002);

\bibitem{SKOT}
T. Shiromizu, K. Koyama, S. Onda and T. Torii, Phys. Rev. {\bf D68}, 063506(2003).

\bibitem{SKT}
T. Shiromizu, K. Koyama and T. Torii, Phys. Rev. {\bf D68}, 103513(2003).

\bibitem{GE}
T. Wiseman, Class. Quant. Grav. {\bf 19}, 3083(2002).

\bibitem{GE2}
S. Kanno and J. Soda, Phys. Rev. D{\bf 66}, 043526(2002);
{\rm ibid}, 083506,(2002).

\bibitem{GE3}
T. Shiromizu and K. Koyama, Phys. Rev. D{\bf 67}, 084022(2003);
S. Kanno and J. Soda, hep-th/0303203, to appear in Gen. Rel. Grav. 

\bibitem{RS}
L.~Randall and R.~Sundrum, Phys. Rev. Lett. {\bf 83}, 4690 (1999).

\bibitem{Tess}
T. Shiromizu, K. Maeda and M. Sasaki, Phys. Rev. {\bf D62}, 024012(2000).


\bibitem{ANT}
A. Aurelia, H. Nicolai and P. K. Townsend, Nucl. Phys. {\bf B176}, 509(1980).

\bibitem{duff}
M. Duff, Phys. Lett. {\bf B226}, 36(1989).

\bibitem{DJ}
M. J. Duncan and L. G. Jensen, Nucl. Phys. {\bf B366}, 100(1990).

\bibitem{holo}
S. S. Gubser, Phys. Rev. {\bf D63}, 084017(2001);
L. Anchordoqui, C. Nunez and L. Olsen, JHEP {\bf 10}, 050(2000);
S. B. Giddings, E. Katz and L. Randall, JHEP {\bf 0003}, 023(2000);
T. Shiromizu and D. Ida, Phys. Rev. {\bf D64}, 044015(2001);
S. de Haro, K. Skenderis and S. N. Solodukhin, hep-th/0011230;
T. Shiromizu, T. Torii and D. Ida, JHEP {\bf 0203}, 007(2002);
S. Nojiri, S. D. Odintsov and S. Zerbini, Phys. Rev. {\bf D62}, 064006(2000);
S. Nojiri and S. D. Odintsov, Phys. Lett. {\bf B484}, 119(2000);
S. W. Hawking, T. Hertog and H. S. Reall, Phys. Rev. {\bf D62}, 043501(2000);
K. Koyama and J. Soda, JHEP {\bf 05}, 027(2001).


\end{thebibliography}
\end{document}